\documentclass[times, 10pt,twocolumn]{article}
\usepackage{latex8}
\usepackage{times}
\usepackage{psfig,url}

\begin{document}

\title{Characterising Web Site Link Structure}

\author{
        Shi Zhou$^*$\\
       {Dept. of Computer Science}\\
       {University College London}\\
       {Adastral Park, Ipswich}\\
       {IP5 3RE, UK}\\
       {s.zhou@ucl.ac.uk}
\and
       Ingemar Cox\\
       Depts. of Computer Science and\\
       Electrical \& Electronic Engineering\\
       {University College London}\\
       {Gower Street, London}\\
       {WC1E 6BT, UK}\\
       {ingemar@ieee.org}
\and
        Vaclav Petricek\\
       {Dept. of Computer Science}\\
       {University College London}\\
       {Gower Street, London}\\
       {WC1E 6BT, UK}\\
       {petricek@acm.org}
}

\maketitle
\thispagestyle{empty}

\begin{abstract}

The topological structures of the Internet and the Web have received considerable attention.
However, there has been little research on the topological properties of individual web sites. In
this paper, we consider whether web sites (as opposed to the entire Web) exhibit structural
similarities. To do so, we exhaustively crawled 18 web sites as diverse as governmental
departments, commercial companies and university departments in different countries.  These web
sites consisted of as little as a few thousand pages to millions of pages.  Statistical
analysis of these 18 sites revealed that the internal link structure of the web sites are
significantly different when measured with first and second-order
topological properties, i.e. properties based on the connectivity of
an individual or a pairs of nodes. 
However, examination of a third-order topological property that
consider the connectivity between three
nodes that form a triangle, revealed a strong correspondence across web sites, suggestive of an
invariant. Comparison with the Web, the AS Internet, and a citation network, showed that this
third-order property is not shared across other types of networks.  Nor is the property exhibited
in generative network models such as that of Barab\'asi and Albert.

\end{abstract} 


\section{Introduction}

The Web has become a global tool for sharing information. It can be
represented as a huge graph which consists of billions of hypertext
web pages connected by hyperlinks pointing from one web page to
another~\cite{border00, levene05a}.  Each web page is part of a larger
web site, which is loosely defined as a group of web pages whose URL
addresses use the same domain name, such as \url{cs.ucl.ac.uk} and
\url{ieee.org}.

Studying and understanding the Web's topological structure is
important as it may lead to improved techniques for information
retrieval. Link structure of the Web has been used in algorithms
like Pagerank~\cite{page98} and HITS~\cite{kleinberg99} to estimate
the importance of web pages, and in \cite{he01,bharat01,kumar99} for
community discovery and clustering. These algorithms do not
typically use the internal link structure within a web site, but
rather, rely on external links between web sites. Nevertheless, the
internal structure of a web site is important.  For example the
statistical property of web site link structure may be used as an
informative measure of web site quality,
e.g.~navigability~\cite{petricek06}.

There is surprisingly little study of the structural properties of web sites {in general}.
Certainly, it is well known that examination of the graph structure of an {individual} web
site can be used to calculate the mean diameter of the web site, and other metrics, that can then
be used to infer properties regarding the navigability of the web site. However, we are unaware of
prior work that provides a {\em statistical} topological
characterization of {all} web sites. As such, web
sites, as opposed to the Web, are often considered to exhibit an
arbitrary statistical topological structure.

However, this study reveals that the topology of web sites is not arbitrary. In fact, examination
of the triangle coefficient (the number of triangles of a node) as a function of degree (the number
of links of the node) reveals a very strong correlation across web sites, suggestive of a possible
invariant of web site link structure.  Moreover, this third-order property varies across other
networks, such as the Web, the Internet and citation networks. Thus, it appears to strongly
characterise web sites.

This paper is organised as follows. In Section~\ref{sec:defs} we
introduce a number of topological metrics which have been used to
characterise and compare network structures. In Section~\ref{sec:data}
we introduce the datasets used in this study. These consist of 18 web
sites which vary in size from a few thousand pages to millions of
pages.  The web sites cover a broad range of entities: 9 government
sites from various countries, 3 commercial sites, 3 educational sites
and 3 very large sites, (IEEE, Wikipedia, Yahoo!).  In
Section~\ref{sec:result} we present our statistical results and
discuss the implications. In addition to comparing data across web
sites, we also compare with (i) subsets of the Web, (ii) a citation
network, (iii) the AS-level Internet network, and (iv) the generative
model of Barab\'asi and Albert~\cite{barabasi99}.
Section~\ref{sec:conc} summarises the key results and discusses how
this work can be used to improve generative models of hypertext
networks.

\section{Definition of Topological Properties}
\label{sec:defs}

We briefly review and define the following topological properties,
which are grouped into three orders according to the scope of
information required to compute them~\cite{Mahadevan06}.
These are (i) the $1^{st}$-order properties, e.g. degree distribution,
(ii) the $2^{nd}$-order properties, e.g.
degree correlation and rich-club connectivity,
and (iii)  the $3^{rd}$-order properties, e.g.
triangle coefficient and clustering coefficient.

\subsection{The $1^{st}$-Order Properties}

\begin{figure}[tbh]
\centerline{\psfig{figure=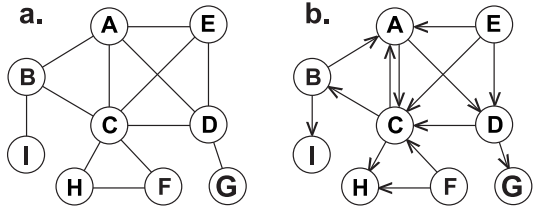,width=7cm}}
\caption{\label{fig:example} Example of (a)~an undirected graph and (b)~a directed graph.}
\end{figure}

The link structure of a web site can be described as an undirected
graph on which a node represents  a web page and a link denotes the
existence of at least one hyperlink connection between two nodes.
The connectivity, or degree $k$, of  a node is defined as the number
of links, or neighbours, the node has. For example in Figure~1a, node
$A$ has four neighbours $B$, $C$, $D$ and $E$, and its degree
$k_A=4$. A web site can also be described as a directed graph on
which each link has a direction pointing from one node to another.
The in-degree $k_{in}$ of a node is then defined as the number of
incoming links and the out-degree $k_{out}$ the number of outgoing
links. For example in Figure~1b, node $A$ has three incoming links
from nodes $B$, $C$ and $E$, i.e.~$k_{in}=3$, and two outgoing links
to nodes  $C$ and $D$, i.e.~$k_{out}=2$.  This paper studies web
sites link structure as undirected graphs unless specifically
stated.

The degree of a node measures a node's local connectivity. Topological
properties calculated by using the degree of individual nodes are
classified as $1^{st}$-order properties, e.g.~the average degree
$\bar{k}$ of nodes in a network.

\subsubsection{Degree Distribution}

The most studied topological property for large networks is the
degree distribution $P(k)$, which is defined as the probability that
a randomly selected node has degree $k$. A random
graph~\cite{erdos60} is characterised by a Poisson degree
distribution where the distribution peaks at the network's average
degree. It has been  reported that a number of
networks~\cite{barabasi99} follow a power-law degree distribution,
\begin{equation}
P(k)\sim k^{-\gamma}, ~<2\gamma<3.
\end{equation}
This means that most nodes have very few links, while a few nodes
have a very large number of links.

\subsection{The $2^{nd}$-Order Properties}

Topological properties are classified as $2^{nd}$-order
properties if they are based on the degree information of the two
end nodes of a link, such as the joint degree distribution
$P(k,k')$~\cite{Dorogovtsev02}, which is the probability that a
randomly selected link connects a node of degree $k$ with a node of
degree $k'$. The $2^{nd}$-order properties provide a more complete
description of a network's structure than the $1^{st}$-order
properties.  For example the degree distribution can be obtained
from the joint degree distribution: $P(k) = (\bar{k}/ k)\sum_{k'}
P(k, k')$.

\subsubsection{Degree Correlation}

The nearest-neighbours
average degree, $k_{nn}$, of  $k$-degree nodes~\cite{Pastor01,
vazquez03}, is a projection of the joint degree distribution, given by
\begin{equation}
k_{nn}(k)=  {\bar{k}\sum_{k'}k'P(k,k')\over kP(k)}.
\end{equation}
A network is called an assortative network if $k_{nn}(k)$ increases
with $k$, which means nodes tend  to attach to similar nodes,
i.e.~high--degree nodes to high--degree nodes and low--degree nodes
to low--degree nodes (`assortative mixing'). Many social networks
are assortative networks. A network is a disassortative network if
$k_{nn}(k)$ decreases with $k$, i.e.~high--degree nodes tend to
connect with low--degree nodes and vice versa (`disassortative
mixing'). This is the case for most information and communications
networks.

A network's degree correlation, or mixing pattern, can be summarised
by a single scalar called the {assortative
coefficient}~\cite{newman02,newman03},
\begin{equation}
\alpha = {L^{-1}\sum_{i} s_i d_i -
[L^{-1}\sum_i{1\over2}(s_i+d_i)]^2 \over
L^{-1}\sum_i{1\over2}(s_i^2+d_i^2)-[L^{-1}\sum_i{1\over2}(s_i+d_i)]^2},
\end{equation}
where $L$ is the number of links and $s_i$, $d_i$ are degrees of the
end nodes of the $i$th link with $i=1,2,...,L$. The value of $\alpha$
is in the range of $[-1, 1]$. For assortative  networks $\alpha>0$
and for disassortative networks $\alpha<0$.

\subsubsection{Rich-Club Connectivity}

The rich-club connectivity~\cite{zhou04a,Colizza06} measures how
tightly the high-degree nodes, \emph{rich} nodes, interconnect
with themselves. If $N_{>k}$ is the number  of nodes with degrees
large than $k$ and they share $E_{>k}$ links between themselves, the
rich-club connectivity is defined as
\begin{equation}
\phi(k) = {2 E_{>k} \over N_{>k}(N_{>k}-1)},
\end{equation}
where $N_{>k}(N_{>k}-1)/2$ is the maximum possible number of links that the $N_{>k}$ nodes can
have. For example in Figure~1a, there are five nodes ($A$, $B$, $C$, $D$ and $E$) with degrees
larger than 2 and they have 8 links between them, thus $\varphi(2)={8\over 5\times(5-1)/2}=0.8$,
which means the 5 best-connected nodes are 80\% fully interconnected. The rich--club connectivity
is a  $2^{nd}$-order property because whether a link belongs to $E_{>k}$ depends on the degrees of
the link's two end nodes.

\begin{figure}[tbh]
\centerline{\psfig{figure=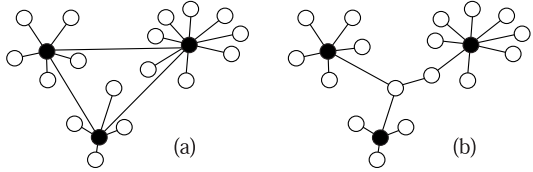,width=7cm}}
\caption{\label{fig:graph-RichClub} Graphs (a) with a rich-club
 and (b) without a rich-club.}
\end{figure}

The rich--club connectivity is a different projection of the joint
degree distribution,
\begin{equation}
\phi(k) = { N\bar{k}\sum_{k',k''=k+1}^{k_{max}}P(k',k'') \over
[N\sum_{k'=k+1}^{k_{max}}P(k')]\cdot[N
\sum_{k'=k+1}^{k_{max}}P(k')-1]},\end{equation} where $N$ is the
total number of nodes and $k_{max}$ is the maximum degree in a
network.  The rich--club connectivity does not trivially relate with
the degree correlation~\cite{zhou07b}. For example the two graphs
shown in Figure~2 are  both disassortative networks, but for the 3
best-connected nodes in Figure~2a, $\phi=1$, and in Figure~2b,
$\phi=0$.

\subsection{The $3^{rd}$-Order Properties}

The $3^{rd}$-order properties are based on connectivity information between three nodes that form a
triangle.

\subsubsection{Triangle Coefficient}

The triangle coefficient $\Delta$ is defined as the number of
triangles a node shares, which is equivalent to the number of links
among the node's neighbours~\cite{Zhou04d}. Triangle is the basic
unit for network redundancy. The more triangles, the more
alternative paths between nodes.

\paragraph{In-triangle and out-triangle coefficients} On a directed
graph, a node's neighbours can be divided into two groups: in-neighbours, which are connected with
incoming links; and out-neighbours, which are connected with outgoing links.  An in-triangle of a
node consists of the node and two of its in-neighbours, and an out-triangle consists of the node
and two out-neighbours. For example in Figure~1b, node $A$ has two in-triangles $ABC$ and $ACE$ and
one out-triangle $ACD$, therefore node $A$'s {in-triangle coefficient} $\Delta_{in}$ is 2 and
{out-triangle coefficient} $\Delta_{out}$ is 1.

\subsubsection{Clustering Coefficient}

A more widely studied $3^{rd}$-order property is the clustering
coefficient $C$, which is defined as the ratio of actual links among
a node's neighbours to the maximal possible number of links they can
share~\cite{watts98}. The clustering coefficient of a node can be
given as a function of a node's degree and its triangle coefficient,
\begin{equation} C={\Delta\over k(k-1)/2}.\label{equation:cluster}
\end{equation}
Two nodes with different triangle coefficients can have  the same
clustering coefficient. For example in Figure~1a, node $B$ has three
neighbours and one triangle and node $C$ has six neighbours and five
triangles ($CBA$, $CAD$, $CAE$, $CED$ and $CFH$). However, their clustering
coefficients are the same:
$$\Delta_B = {1\over 3(3-1)/2}={1\over 3}={5\over  6(6-1)/2}=\Delta_C.$$
Therefore one should use the triangle
coefficient to infer the clustering information of nodes with
different degrees.

\section{Datasets}
\label{sec:data}

Here we briefly summarise the various datasets used in this study.

\begin{table*}
\centering \caption{Properties Of The Datasets} \label{table:websites}
\centering
\begin{tabular}{|l|r|r|r|r|r|r|}
\hline
                        &    Web Site    &   Number &    Number  &   Average        &   Assortative &   Average     \\
    Dataset     &    domain name &   of nodes    &   of links    &   degree   &    coefficient  &triangle  coef. \\
\hline  AO-CA   &   cac.gc.ca       &    12,730     &    120,485    &    15.94  &     -0.35  &    159.78     \\
  AO-IT   &   corteconti.it   &    32,614     &    200,516    &    11.96  &    -0.40   &    186.11     \\
  AO-UK   &   nao.gov.uk      &    4,027  &    25,453     &    11.84   &     -0.36  &    89.40  \\
  AO-US   &   gao.gov         &    19,625     &    223,998    &    21.69  &    -0.63   &    289.37     \\
\hline  FO-AU   &   dfat.gov.au     &    29,140     &    791,039    &    53.25  &   -0.78    &    1,066.30   \\
  FO-CZ   &   mzv.cz          &    31,246     &    778,163    &    45.23  &   -0.13    &    1,134.06   \\
  FO-DE   &   auswaertiges-amt.de &    46,219     &    2,234,535  &    94.10  &   -0.56    &    4,439.89   \\
  FO-JP   &   mofa.go.jp      &    52,206     &    493,861    &    17.11  &   -0.37    &    177.23     \\
  FO-UK   &   fco.gov.uk      &    33,280     &    694,255    &    36.29  &   -0.16    &    884.54     \\
\hline  COM-HSBC    &   hsbc.co.uk  &    51,043     &    68,454     &    2.62   &   -0.05    &    7.97   \\
  COM-NEXT    &   next.co.uk  &    74,989     &    557,466    &    14.11  &   -0.47    &    182.55     \\
  COM-SKODA   &   skoda-auto.com  &    49,341     &    727,119    &    28.39  &    -0.30   &    292.12     \\
\hline  EDU-AUCK    &   arts.auckland.ac.nz &    12,457     &    129,870    &    17.64  &  -0.21     &    258.13     \\
  EDU-UCB &   haas.berkeley.edu   &    100,025    &    373,521    &    6.90   &    -0.09   &    84.85  \\
  EDU-UCL &   cs.ucl.ac.uk    &    36,554     &    229,711    &    10.81  &   -0.15    &    70.34  \\
\hline  LARGE-IEEE  &   ieee.org    &    1,977,923  &    5,614,610  &    5.54   &   -0.05    &    57.92  \\
  LARGE-WIKI  &   zh.wikipedia.org    &    1,913,510  &    8,249,248  &    8.12   &    -0.13   &    64.54  \\
  LARGE-YAHOO &   yahoo.com   &    3,448,289  &    12,039,165     &    6.72   &   -0.08    &    81.69  \\
\hline  Web   &     --  &    43,425    &   173,696     &   7.96    &   -0.12    &    38.43   \\
Citation network    &     --  &    244,864    &    897,170    &   7.33    &   -0.08    &    4.20   \\
AS Internet    &     --  &    9,200    &    28,957    &   6.30   &    -0.24   &  21.37     \\
 BA model    &             --  &    10,000     &   30,000  &   6.00   &    -0.02   &   0.16    \\
\hline\end{tabular}
\end{table*}

\subsection{Web sites}

We exhaustively crawled the 18 web sites of the organisations
listed in Table 1: 1)~the national audit office or equivalent of
Canada (AO-CA), Italy (AO-IT), the United Kingdom (AO-UK) and the
United States (AO-US); 2)~the foreign office or equivalent of
Australia (FO-AU), the Czech Republic (FO-CZ), German (FO-DE), Japan
(FO-JP) and the UK (FO-UK); 3)~commercial web sites, such as HSBC
bank in the UK (COM-HSBC), the UK retailer NEXT (COM-NEXT) and the
automobile company SKODA (COM-SKODA); 4)~educational web sites, such
as the Faculty of Arts at the University of Auckland, New Zealand (EDU-AUCK),
the Haas School of Business at the University of California at Berkeley
(EDU-UCB), and the Department of Computer Science at University College
London (EDU-UCL); and 5)~three very large web sites with millions of
web pages, such as the IEEE (LARGE-IEEE), Wikipedia in the language of Simplified
Chinese (LARGE-WIKI) and Yahoo! (LARGE-YAHOO).

We used the Nutch 1.6.0 crawler
(\url{http://lucene.apache.org/nutch}). Each crawl was started from a
web site's homepage and was restricted to the web site's domain as
listed in Table~1. The crawler was configured to allow for complete
site acquisition and collected all web pages up to a depth of
18. The default parameters were a 5-second delay between requests to
the same host, and 10,000 attempts to retrieve pages that fail with a
`soft' error~\cite{petricek06}. We discarded hyperlinks pointing to
web pages outside the web site's domain and removed self-loops and
duplicated hyperlinks.

We are aware of a number of available data sources of the Web. We
did not extract web sites data from them because they aim to sample
the entire Web and contain very incomplete information of the
internal link structure of individual web sites. For example the
Stanford WebBase data
(\url{http://dbpubs.stanford.edu:8091/~testbed/doc2/WebBase})
contains only 400 web pages with NASA's domain name (\url{nasa.org}).

\subsection{Web}

WT10g is a mega dataset of the Web proposed by the annual
international Text REtrieval Conference (TRECs,
\url{http://trec.nist.gov}). WT10g is constructed from more than 320
gigabytes of archived data containing 1.7M  web pages and hyperlinks
between them. It is reported that WT10g retains properties of the
larger Web~\cite{Soboroff02} and has been used as a data resource
for research on Web retrieval and modelling. We randomly sampled
10 subsets of WT10g, each of which contains 50,000 web pages and
links between those pages. In this paper we use the average
properties of the 10 WT10g subsets as an approximation of the Web's
link structure.

\subsection{Citation Network}

The citation network~\cite{petricek05} data was extracted from the
online computer science publication database CiteSeer
(\url{http://citeseer.ist.psu.edu/}). The CiteSeer data contain 575K
entries, from which we extracted 244,864 records having at least one
reference (outgoing link) or citation (incoming link).

\subsection{AS Internet}

The Internet topology at the autonomous systems (AS) level has been
extensively studied in recent years~\cite{Pastor04, Zhou04d,
  mahadevan05b, Mahadevan06}. On the AS Internet, nodes represent
Internet service providers and links represent connections between
them. In this paper we use the AS Internet dataset ITDK0304 collected
by CAIDA~\cite{CAIDA}.

\subsection{Barab\'asi-Albert Model}

The Barab\'asi and Albert
(BA) model~\cite{barabasi99} has been widely used in the study
of complex networks. This model shows  that a power-law
degree distribution can be produced by two mechanisms: \emph{growth}, where the network ``grows'' from
a small random graph by attaching new nodes to old nodes in the
existing system; and \emph{preferential attachment}, where a new
node is attached preferentially to nodes that are already well
connected.

\section{Results}
\label{sec:result}

Here we summarise our experimental findings. We examine a variety of first, second and third-order
topological properties and compare them across the various web sites.  We then compare the
topological properties of web sites with other networks, specifically, the Web, AS network, a
citation network, and the generative network of Barab\'asi and Albert.

\subsection{Comparison between the web sites}

\begin{figure*}[tbh]
\centerline{\psfig{figure=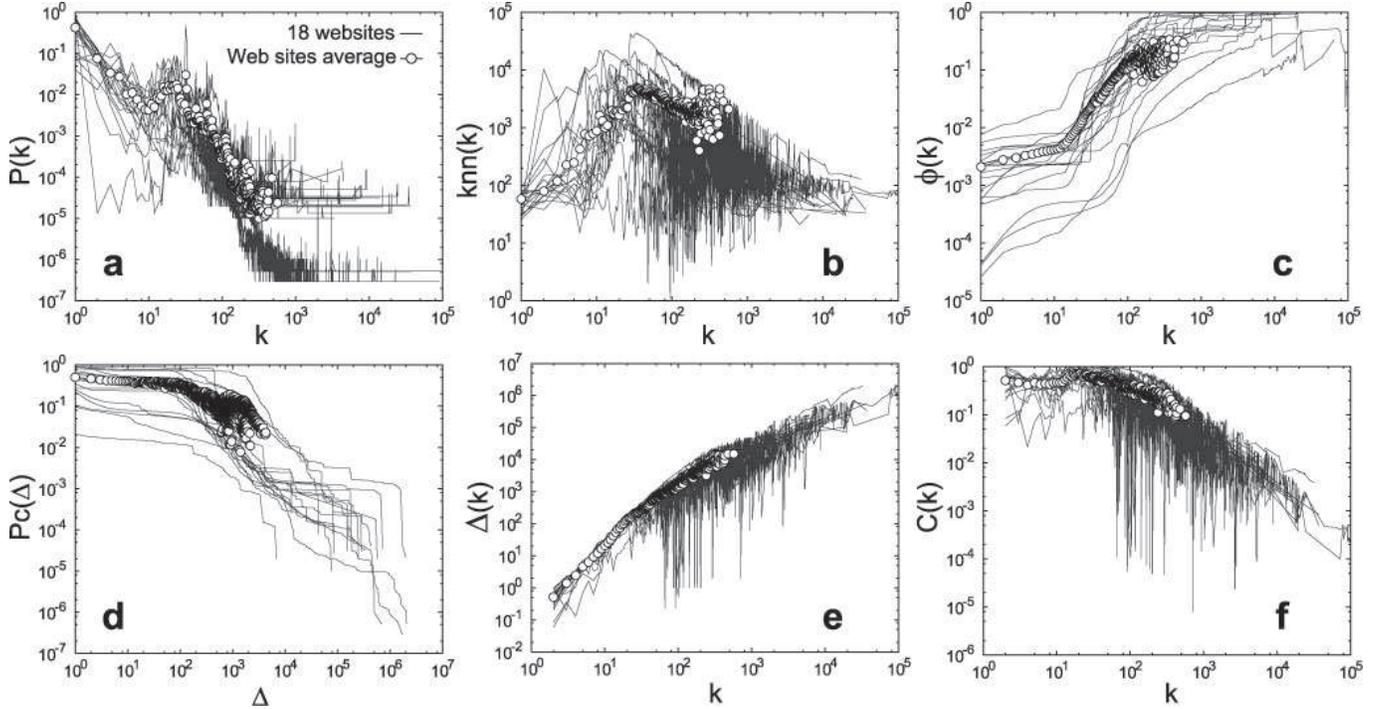,width=18cm}}
\caption{\label{fig:websites} Topological properties of the
web sites: a)~degree  distribution, $P(k)$; b)~nearest-neighbours
average degree of $k$-degree nodes, $k_{nn}(k)$; c)~rich-club
connectivity as a function of degree, $\phi(k)$; d)~complementary
cumulative distribution of triangle coefficient, $P_{c}(\Delta)$;
e)~correlation between triangle coefficient and degree, $\Delta(k)$;
and f)~correlation between clustering coefficient and degree,
$C(k)$. }
\end{figure*}

\subsubsection{The $1^{st}$ And $2^{nd}$-Order Properties}
As shown in Table~1, the size and the average degree of the web sites vary significantly.  The
foreign office web sites have very large average degrees, whereas the three large web sites with
millions of web pages have very small average degrees. Figure~\ref{fig:websites}a, b~and~c
illustrate the degree distribution $P(k)$, the degree correlation $k_{nn}(k)$,  and the rich-club
connectivity $\phi(k)$ of the 18 web sites on a log-log scale. Also shown are their \emph{average}
properties, depicted by circles\footnote{ The average degree distribution $\bar{P}(k)$ is obtained
as such: for a given $k$, if at least $X\ge 12$ of the 18 web sites have $P(k)>0$, then $\bar{P}(k)
= X^{-1} \sum_i P_i(k)$ where $i=1,2...X$. Other average properties are calculated in similar ways.
}. It is clear that the $1^{st}$ and $2^{nd}$-order properties of the web sites exhibit huge
variations over several orders of magnitudes.  Thus, the web sites cannot be well characterised by
the average of these properties. For example, in Figure~\ref{fig:websites}c, some web sites with
nodes of degree $k>100$ are almost fully interconnected with themselves, i.e.~$\phi\approx 1$,
whereas in other web sites the interconnectedness is much looser, with $\phi$ less than 0.001.

\subsubsection{The $3^{rd}$-Order Properties}
\label{sec:thirdorder}

Figure~\ref{fig:websites}d shows the complementary cumulative
distribution of the triangle coefficient $P_c(\Delta)$, which is the
probability that a node's triangle coefficient is larger than
$\Delta$. Figure~\ref{fig:websites}e shows the relationship between
triangle coefficient  and degree $\Delta(k)$, i.e.~the average
triangle coefficient of $k$-degree nodes. Although the web sites do
not show an agreement on $P_c(\Delta)$, they do exhibit a clear
correspondence on $\Delta(k)$. Some web sites have sharp spikes on
their $\Delta(k)$ curves. These spikes reflect the existence of
star-like subgraphs in these web sites, e.g.~a web page with a long
list of hyperlinks pointing to documents or images. Compared to
the large number of web pages contained in a web site, the limited
number of such spikes are not statistically significant.

The {\em average} over all the web sites of the triangle coefficient
as a function of degree is also depicted in
Figure~\ref{fig:websites}e, see circles, and
is a smooth curve, which well represents all the
web sites.  This is suggestive of a structural invariant of web
sites.

Figure~\ref{fig:websites}f shows  the web sites show a similar
correspondence  on the relationship between clustering coefficient
and degree $C(k)$.  Note that the {\em average} clustering
coefficient, depicted by circles, is not a
monotonic function of degree.  This is because the clustering coefficient
is itself a function of the degree and triangle coefficient. In the
following we do not consider $C(k)$ further, as the triangle coefficient,
$\Delta(k)$, contains all information provided by $C(k)$.

\subsection{Comparison with other networks}

Here we compare the topological properties of the {\em average} over
all web sites, with those of other networks, specifically the Web, a
citation network, the AS Internet, and the BA model.

\begin{figure*}[tbh]
\centerline{\psfig{figure=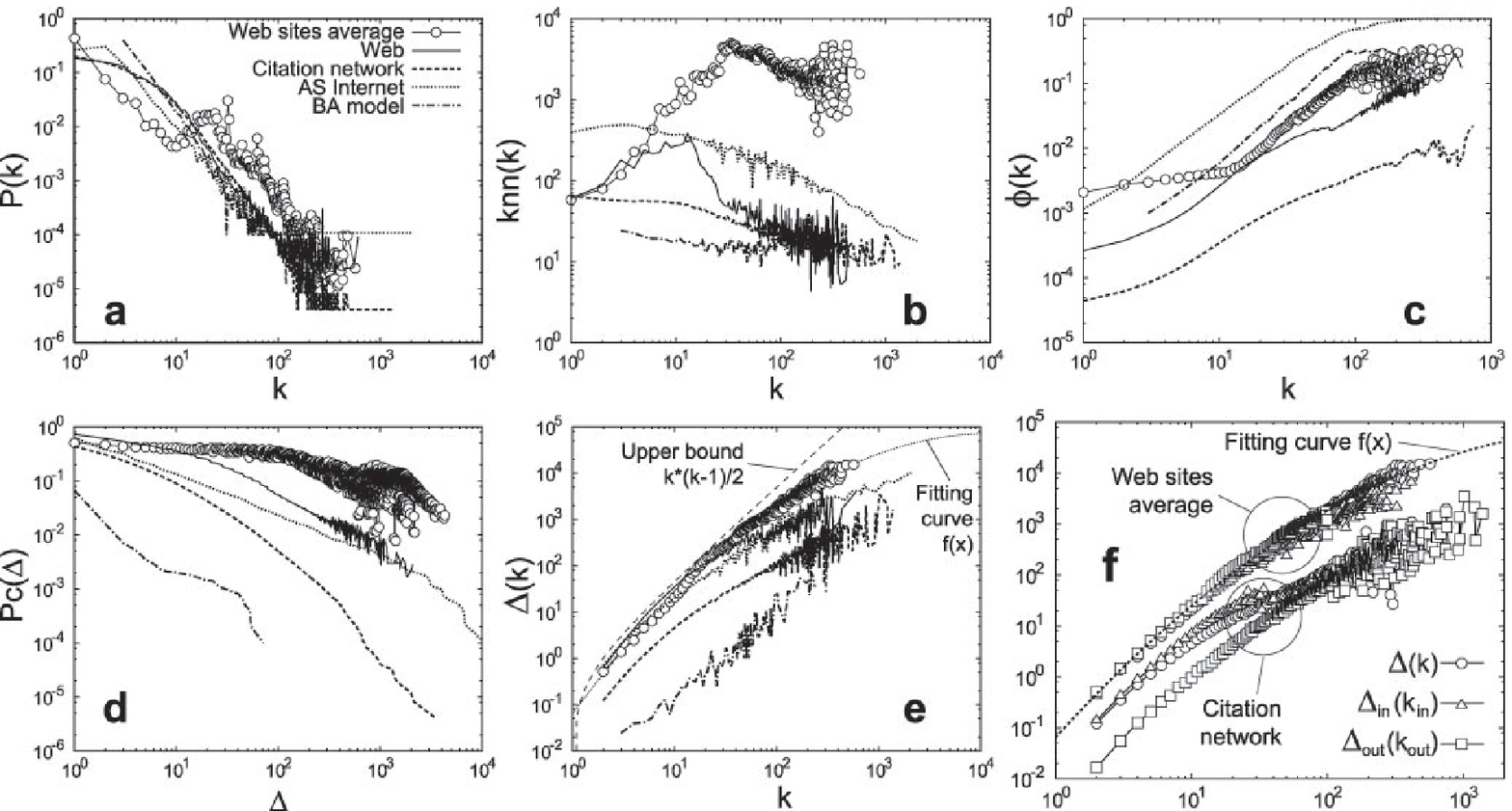,width=18cm}} \caption{\label{fig:networks} Comparison
between the average of the web sites and (i) the  Web, (ii) a citation
network, (iii) the AS Internet, and (iv) the BA model:
a)~degree distribution, $P(k)$; b)~nearest-neighbours average degree of $k$-degree nodes,
$k_{nn}(k)$; c)~rich-club connectivity as a function of degree, $\phi(k)$; d)~complementary
cumulative distribution of triangle coefficient, $P_{c}(\Delta)$; e)~  triangle
coefficient as a function of degree, $\Delta(k)$; and f)~three triangle properties: triangle coefficient versus
degree,~$\Delta(k)$; in-triangle coefficient versus in-degree,~$\Delta_{in}(k_{in})$; and out-triangle
coefficient versus out-degree,~$\Delta_{out}(k_{out})$. }
\end{figure*}

\subsubsection{Degree Distribution}

Figure~\ref{fig:networks}a shows that the degree distribution  of
the Web, the citation network, the AS Internet and the BA model can
be well described as a power-law $P(k)\sim k^{-\gamma}$ with
$2<\gamma<3$. However the average degree distribution of the
web sites is very different: for $k<10$ or $k>30$, it can be
described as a power-law; but for $10<k<30$,  the distribution
increases exponentially with degree.

\subsubsection{Degree Correlation}

Figure~\ref{fig:networks}b shows that the citation network and the AS
Internet are typical disassortative networks where $k_{nn}$
decreases monotonically with $k$. The BA model is an example of a
neutral network where $k_{nn}$ does not change with $k$. For the
average of the
web sites, and the Web, $k_{nn}$ first increases and then decreases
with $k$, and peaks at $k=30$ and $k=15$ respectively. For large
degrees, the average $k_{nn}$ of the web sites is significantly
larger than all other networks.

\subsubsection{Rich-Club Connectivity}

Figure~\ref{fig:networks}c shows that the AS Internet has the highest
rich-club connectivity, with a fully interconnected core, i.e. $\phi(k)=1$, for $k>200$. The
citation network has the lowest rich-club connectivity. Although the BA model is very different
from the web sites when measured by $k_{nn}(k)$, the two exhibits
similar rich-club connectivity for $k>10$.

\subsubsection{Distribution of Triangle Coefficient}

Figure~\ref{fig:networks}d shows that the web sites contain significantly more triangles than all
other networks.  The high density of triangles ensures the navigability of the web sites.

\subsubsection{Triangle Coefficient as a Function of Degree}

Figure~\ref{fig:networks}e shows that, in general, all the networks exhibit a positive correlation
between triangle coefficient and degree.  This is because the larger the degree of a node, the more
neighbours a node has, and thus the higher the chance of forming triangles.  As discussed in
Section~\ref{sec:thirdorder},  all the web sites exhibit a very similar relationship between
triangle coefficient and degree, that is well characterised by the average over all the web sites.
The average correlation between triangle coefficient and degree of the web sites can be closely
fitted by a function given as
$$f(x)=0.064x^{2.94-0.36\log_{10}(x)}$$ or $$\log_{10}(f(x))=-0.3579\log_{10}^2(x)+
2.9432\log_{10}(x)-1.1907.$$
It is clear that the relationship between triangle coefficient and degree is different
from the other networks.  The BA model exhibits the lowest number of
triangles as a function of node degree, followed by the citation
network, and then the AS Internet.  For degree $k<30$, the Web data closely follows that
of the average over web sites, but diverges thereafter.

\subsubsection{In-Triangle and Out-Triangle}

Figure~\ref{fig:networks}f examines the three relationships of (i) triangle
coefficient versus degree $\Delta(k)$, (ii)
in-triangle coefficient versus in-degree $\Delta_{in}(k_{in})$, and (iii)
out-triangle coefficient versus out-degree
$\Delta_{out}(k_{out})$, for the citation network and the average over
all 18 web sites.  That is, here, we consider the networks as {\em
directed} graphs.

For the web sites, these three relationships closely overlap one
another.  This means the probability of forming triangles with a web
page's in-neighbours or with its out-neighbours are the same.
However, for the
citation network, $\Delta_{in}(k_{in})$ is one order of
magnitude larger than $\Delta_{out}(k_{out})$ for the same
degrees. This means the probability of a paper forming triangles with
its citations (in-neighbours) is significantly larger than it forming
triangles with its references (out-neighbours).

This structural difference between web sites and the citation network
may reflect their different evolution dynamics. For a citation
network, when a paper is published all its references existed before
the publication of the paper and, of course, cannot be changed.
However, a paper can always acquire new citations, and these citations
may reference other citations (thus continuing to form triangles).  In
contrast, for a web site, web pages and their associated hyperlinks
can be added, deleted or revised at any time.  For web sites, there is
no equivalent to a reference to a page that remains static and unable
to be changed in the future.

\section{Conclusion}
\label{sec:conc}

We examined a number of topological properties of hyperlink data
crawled from 18 diverse web sites.  Our empirical results showed that
the link structures of the web sites are significantly different when
measured with 1st and 2nd-order topological properties. {This
  is probably to be expected since the web sites are designed for
  different purposes and developed independently.} However we observed
that web sites share a common 3rd-order topological property, the
relationship between triangle coefficient and degree. This common
relationship is unexpected and suggestive of a topological invariant
for web sites. Comparison with the Web, the AS Internet, a citation
network and the Barab\'asi-Albert model showed that this third-order
property is not shared across other types of networks. Thus, this
property appears to strongly characterise web sites. { The
  physical meaning of this 3rd-order property is that given the
  number of hyperlinks to and from a particular web page, we can {\em
    statistically} estimate how the web page's neighbouring pages are
  interlinked; and this statistical estimation is valid for all web
  sites.} 

{Further evaluation on a wider variety of web sites is needed
  to verify that this 3rd-order property is an invariant. If so, then
  the fundamental question is why? Possible explanations include
  standardised web site designing principles, popular web site
  developing tools, or universal evolution dynamics which
  fundamentally reflect the common nature and function of web sites as
  a way of organising and disseminating information. The answer to
  this question may prove valuable for research on a number of issues,
  such as modelling web site and other document networks,
  recommendations for building web sites in the future, optimizing
  search engine algorithms, and understanding the fundamental
  principles governing the evolution of the Web.}

\section{Acknowledgments}

This work is partly supported by the UK Nuffield Foundation grant
NAL/01125/G and a grant from the Cambridge-MIT Institute.


\begin{thebibliography}{50}\setlength{\itemsep}{-1ex}\small

\bibitem{CAIDA}
{The Cooperative Association For Internet Data Analysis}.
\newblock \url{http://www.caida.org/}.

\bibitem{barabasi99}
A.~Barab\'asi and R.~Albert.
\newblock Emergence of scaling in random networks.
\newblock {\em Science}, 286:509, 1999.

\bibitem{bharat01}
K.~Bharat, B.-W. Chang, M.~Henzinger, and M.~Ruhl.
\newblock Who links to whom: Mining linkage between web sites.
\newblock In {\em Proc. of IEEE Intl. Conf. on Data Mining (ICDM)}, 2001.

\bibitem{border00}
A.~Broder, R.~Kumar, F.~Maghoul, S.~R. P.~RaghavanRajagopalan, S., and
  A.~Tomkins.
\newblock Graph structure in theweb: Experiments and models.
\newblock In {\em WWW'00: Proc.of the 9th Intl. Conf. on World Wide Web}, 2000.

\bibitem{Colizza06}
V.~Colizza, A.~Flammini, M.~A. Serrano, and A.~Vespignani.
\newblock Detecting rich-club ordering in complex networks.
\newblock {\em Nature Physics}, 2:110--115, 2006.

\bibitem{Dorogovtsev02}
S.~N. Dorogovtsev and J.~F.~F. Mendes.
\newblock Evolution of networks.
\newblock {\em Adv. Phys.}, 51(1079), 2002.

\bibitem{erdos60}
P.~Erd\H{o}s and A.~R\'enyi.
\newblock On the evolution of random graphs.
\newblock {\em Publ. Math. Inst. Hung. Acad. Sci.}, 5:17, 1960.

\bibitem{he01}
X.~He, H.~Zha, C.~Ding, and H.~Simon.
\newblock Web document clustering using hyperlink structures.
\newblock {\em Computational Statistics and Data Analysis}, 41(1):19--45, 2001.

\bibitem{kleinberg99}
J.~Kleinberg.
\newblock Authoritative sources in a hyperlinked environment.
\newblock {\em Journal of the ACM}, 46(5):604--632, 1999.

\bibitem{kumar99}
R.~Kumar, P.~Raghavan, S.~Rajagopalan, and A.~Tomkins.
\newblock Trawling the web for emerging cyber-communities.
\newblock {\em Computer Networks}, 31(11-16):1481--1493, 1999.

\bibitem{levene05a}
M.~Levene.
\newblock {\em An Introduction to Search Engines and Web Navigation}.
\newblock Pearson Education, 2005.

\bibitem{Mahadevan06}
P.~Mahadevan, D.~Krioukov, K.~Fall, and A.~Vahdat.
\newblock Systematic topology analysis and generation using degree
  correlations.
\newblock In {\em Proc.~of SIGCOMM'06}, pages 135--146. ACM Press, New York,
  2006.

\bibitem{mahadevan05b}
P.~Mahadevan, D.~Krioukov, M.~Fomenkov, B.~Huffaker, X.~Dimitropoulos,
  K.~Claffy, and A.~Vahdat.
\newblock {The Internet AS-level Topology: Three Data Sources and One
  Definitive Metric}.
\newblock {\em Comput. Commun. Rev.}, 36(1):17--26, 2006.

\bibitem{newman02}
M.~E.~J. Newman.
\newblock Assortative mixing in networks.
\newblock {\em Phys. Rev. Lett.}, 89(208701), 2002.

\bibitem{newman03}
M.~E.~J. Newman.
\newblock Mixing patterns in networks.
\newblock {\em Phys. Rev. E}, 67(026126), 2003.

\bibitem{page98}
L.~Page, S.~Brin, R.~Motwani, and T.~Winograd.
\newblock The pagerank citation ranking: Bringing order to theweb.
\newblock Technical report, Stanford Digital Library Technologies Project,
  1998.

\bibitem{Pastor01}
R.~Pastor-Satorras, A.~V\'azquez, and A.~Vespignani.
\newblock Dynamical and correlation properties of the internet.
\newblock {\em Phys. Rev. Lett.}, 87(258701), 2001.

\bibitem{Pastor04}
R.~Pastor-Satorras and A.~Vespignani.
\newblock {\em Evolution and Structure of the Internet - A Statistical Physics
  Approach}.
\newblock Cambridge University Press, Cambridge, 2004.

\bibitem{petricek05}
V.~Petricek, I.~J. Cox, H.~Han, I.~Councill, and C.~L. Giles.
\newblock A comparison of on-line computer science citation databases.
\newblock In {\em ECDL'2005: Proc. of the 9th European Conf. on Research and
  Advanced Technology for Digital Libraries}. Springer, 2005.

\bibitem{petricek06}
V.~Petricek, T.~Escher, I.~J. Cox, and H.~Margetts.
\newblock The web structure of e-government - developing a methodology for
  quantitative evaluation.
\newblock In {\em WWW'06: Proc.of the 15th Intl. Conf. on World Wide Web},
  2006.



\bibitem{Soboroff02}
I.~Soboroff.
\newblock Does wt10g look like the web?
\newblock In {\em ACM SIGIR'02}, pages 423--425, 2002.

\bibitem{vazquez03}
A.~V\'azquez, M.~{Bogu\~n\'a}, Y.~Moreno, R.~Pastor-Satorras, and
  A.~Vespignani.
\newblock Topology and correlations in structured scale-free networks.
\newblock {\em Phys. Rev. E}, 67(046111), 2003.

\bibitem{watts98}
D.~J. Watts and S.~H. Strogatz.
\newblock Collective dynamics of `small-world' networks.
\newblock {\em Nature}, 393:440, 1998.

\bibitem{zhou07b}
S.~Zhou and R.~Mondrag\'on.
\newblock Structural constraints in complex networks.
\newblock {\em New J. of Physics}, 9(173), 2007.

\bibitem{Zhou04d}
S.~Zhou and R.~J. Mondrag\'on.
\newblock Accurately modelling the {I}nternet topology.
\newblock {\em Phys. Rev. E}, 70(066108), 2004.

\bibitem{zhou04a}
S.~Zhou and R.~J. Mondrag\'on.
\newblock The rich-club phenomenon in the {I}nternet topology.
\newblock {\em IEEE Comm. Lett.}, 8(3):180--182, 2004.





\end{thebibliography}

\end{document}